\documentstyle[prc,aps,psfig]{revtex}
\newcommand{\be}{\begin{eqnarray}}
\newcommand{\ee}{\end{eqnarray}}

\newcommand{\bi}{\begin{itemize}}
\newcommand{\ei}{\end{itemize}}
\begin{document}
\twocolumn[\hsize\textwidth\columnwidth\hsize
           \csname @twocolumnfalse\endcsname
\title{Chaotic scattering on surfaces and collisional damping of collective
modes}
\author{Klaus Morawetz$^{1,2,3}$, Michael Vogt$^4$, Uwe Fuhrmann$^3$, Pavel Lipavsk\'y$^5$, V\'aclav \v Spi\v cka$^5$}
\address{$^1$ LPC, IN2P3-CNRS, ISMRA, 6 Bd du Marechal Juin, Caen, 14050, France\\
$^2$ GANIL,
B.P. 5027,
Bd Henri Becquerel,
14076 Caen Cedex 5, France\\
$^3$ Fachbereich Physik, University Rostock, D-18055 Rostock,
Germany\\
$^4$ Institute of Philosophy, University Rostock, D-18055 Rostock,
Germany\\
$^5$ Institute of Physics, Academy of Sciences, Cukrovarnick\'a 10,
16200 Praha 6, Czech Republic}
\maketitle
\begin{abstract}

The damping of hot giant dipole resonances is investigated. The contribution of surface scattering is compared with the contribution from interparticle 
collisions. A unified response function is presented which includes surface damping as 
well as collisional damping. The surface damping enters the response via the Lyapunov 
exponent and the collisional damping via the relaxation time. The former is 
calculated for different shape deformations of quadrupole and octupole type. The 
surface as well as the collisional contribution each reproduce almost the experimental 
value, therefore we propose a proper weighting between both contributions related to their 
relative occurrence due to collision frequencies between particles and of particles with 
the surface. We find that for low and high temperatures the collisional contribution 
dominates whereas the surface damping is dominant around the temperatures $\sqrt{3}/2\pi$ of the centroid energy. 
\end{abstract}
\vskip2pc]

\section{Introduction}

The damping mechanisms of collective motions in excited nuclei
are a topic of continuing debate \cite{BBB83}. 
Mainly two lines of thought are pursued.
In one line of thought it is assumed that collisions are the physical reason for damping only as developed via a Fermi liquid approach with  bulk matter
properties 
\cite{KAM69,SAB84,ATS94,KPS95,KPS96,KTO96,HNP96,TKL97,KLT97,FMW98,KLP98,MWF97,AYG98,GTO98}.
The other line of thought considers new features of the finite nucleus,    
such as
surface oscillations and a level
density with finite spacing.  Partially the investigations are performed without inertia \cite{SHB75,ESB83,ESB84,BBB91,OBB96,KST97,OBB97,KLL98,KIN98}
or by including inertia \cite{OBB97,ALB89,OBB90,OCB92,LBB95,BCM95,LDH96}; note that inertia is absent in infinite matter. 

Both classes of models predict a comparable degree of damping necessary
to reproduce the experimental data. Consequently, it is an open question which is the correct physical reason for damping.
Of course, the correct
description has to assume a finite nucleus consisting of nucleons which are
bound via the mean field, through which the nucleons undergo mutual collisions and where the surface is formed by the particles themselves.
These features are usually included in Boltzmann-Uehling-Uhlenbeck-(BUU) simulations 
\cite{CTG93,BAR96,MWS98} or in its nonlocal extensions \cite{SLM96,MLSCN98}. In full simulations, however,  we will not gain a simple insight into the physical origins of the damping mechanism, in particular, how much is due to surface contributions and how much is due to collisional contributions.

The aim of this article is to compare both pictures
in the frame of linear response theory. 
Within the collision-free Vlasov equation the linear response of finite systems is well known \cite{BDT86} and allows one to calculate the strength function of finite nuclei. The damping, however, does not reproduce the experimental damping of giant resonances since collisions are absent.

To include damping, we can take into account on the one hand 
the collisional damping in infinite matter and find a scaling for 
finite size effects in the sense of a Thomas-Fermi local density approximation. On the 
other hand
we can consider the boundary of the finite nucleus as 
a fixed surface such that the nucleons simply bounce
off this wall like in a billiard-type situation. Provided we accept this simple picture, it is possible to
compare the damping caused by this chaotic
scattering off the wall with the damping from collisional
contributions in infinite matter. We will derive a response function which includes the additional 
chaotic process and find a total damping rate $\Gamma=\Gamma_{\rm coll}
+\Gamma_{\rm surf}$ similar 
to the Matthiessen rule in metals \cite{MJ36} but with an additional weight between the different processes according to the ratio of the corresponding collision frequencies.
Moreover the hypotheses by Swiatecki \cite{swiatecki} is questioned that negative curved surfaces of octupole deformations would induce an additional chaotic mechanism of dissipation and consequently octupole modes should be overdamped.
We will find that the  octupole deformation does not lead to any special
enhancement of the Lyapunov exponent in comparison with the quadrupole mode. 
Moreover, any deformation will cause a contribution to the damping of collective modes.

The outline of the paper is as follows. In the next chapter \ref{II} the largest Lyapunov exponent for a nucleon in a deformed nucleus is calculated. Then in chapter \ref{III} we derive a unified response function which combines both the collisional damping and the contribution from surface scattering. In chapter \ref{IV} both damping contributions are compared with the experimental values of hot isovector giant dipole resonances (IVGDR). A proper weighting factor corresponding to the relative collision frequencies yields a unified picture which describes the data rather well. We find that the collisional contributions dominate for low temperatures and for very high temperatures while around the highest experimental achievable temperatures the surface effects are important. Chapter \ref{V} will summarize the results.

\section{Largest Lyapunov exponent of deformed nuclei}\label{II}

We will consider in the following only classical 3-D closed billiards and use the largest Lyapunov exponent as a relevant measure to characterize chaoticity. The surface of these billiards is chosen to resemble 
the surface deformation of a nucleus undergoing quadrupole or octupole deformations.
 
The Lyapunov exponent can be given as
the deviations of the difference of the trajectories 
\be
|\vec{r_1}(t)-\vec{r_2}(t)| =  |\vec{r_1}(0)-\vec{r_2}(0)| \, 
e^{\lambda t}
\ee
as
\be
\lambda & = & \lim_{t\to\infty} \, \lim_{\varepsilon\to 0} \, \frac{1}{t}\,
\ln \frac{|\vec{r_1}(t)-\vec{r_2}(t)|}{|\vec{r_1}(0)-\vec{r_2}(0)|}.
\ee
For $\lambda > 0$ the difference in phase space trajectories grows exponentially leading to
chaotic behavior.

We solve the Hamilton equations for
one particle in finite nuclear matter
with an infinite potential\footnote{The numerical implementation is performed with [$V_0\rightarrow \infty$ and $\sigma\rightarrow 0$]
\[
V(r,\theta)\approx V_0 \,\left(\arctan\, 
                           \left(\frac{r-R(\theta)}{\sigma}\right)+{\pi \over2}\right )
\]
resembling an infinite step function.}
\be
V(r,\theta)=V_0 \,\Theta\, \left(r-R(\theta)\right) 
\ee
modeling the deformation of a axial symmetric  nuclear
surface 
\be
R_{\lambda}(\theta)
=R_0\,\left(1+\alpha_{00}+
             \alpha_{\lambda}\,P_{\lambda}(\cos{(\theta)})\right)\label{r1}
\ee
with the nuclear radius $R_0=1.13 A^{1/3}$fm, and where
$\lambda=2$ corresponds to the quadrupole and 
$\lambda=3$ to the octupole deformation \cite{ringschuck}.
The coefficient $\alpha_{00}$ is adjusted to conserve the volume corresponding to incompressible nuclear matter.

\begin{figure*}[h]
\centerline{\psfig{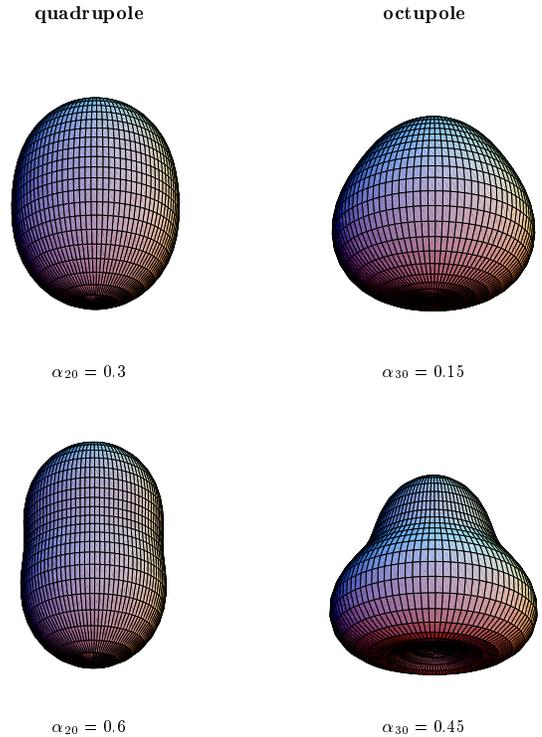}}
\caption{Overview of the geometry of modes}
\end{figure*}

The Lyapunov exponent is calculated by considering the time evolution of 
small deviations from a reference 
trajectory due to infinitesimal initial changes.
We have used the Brandstaetter method resetting the deviations of the reference
trajectory repeatedly after a certain time to the initial infinitesimal difference. This corresponds to an 
averaging and the largest mean Lyapunov exponent is obtained.

\begin{figure}[t]
  \centerline{\psfig{file=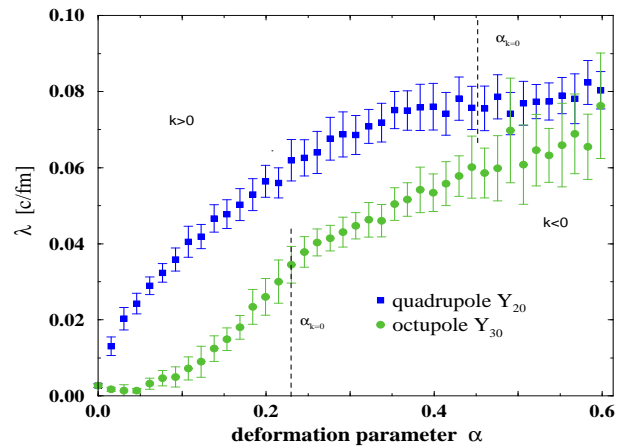,width=8cm,height=6cm,angle=-90}}
\vspace{.5cm}
\caption{\label{fig1} The largest Lyapunov exponent of a spherical deformed 
billiard versus deformation parameter 
(\protect\ref{r1}). The regions of $\alpha$ where the surface starts to become negatively 
curved ($k<0$) via (\protect\ref{k}) are indicated by dotted lines $\alpha_{k=0}$ for octupole and quadrupole deformation, respectively.
}
\end{figure}

Figure \ref{fig1} shows the Lyapunov exponent for different 
octupole and quadrupole 
deformations according to (\ref{r1}). The error bars are taken from averaging 50 runs of 
different initial conditions indicating $95\%$ 
confidence level. We see that the quadrupole deformation leads to 
an immediate increase of the Lyapunov exponent while the Lyapunov exponent for the
octupole deformation increases at larger deformation parameters and proceeds more slowly with increasing deformation. 
Let us note that the characteristic time scales from the Lyapunov exponent are similar to the damping rate of the IVGDR which will be presented in 
figure \ref{fig4}. This motivates us to consider a unified response from collisions and surface contributions.

Following \cite{swiatecki} a surface with negative curvature should induce a new chaotic mechanism analogous to the Sinai - billiard. We discuss therefore the curvature of a rotating body given
by the curvature of the boundary curve $R(\theta)$ of (\ref{r1}) via
\be
k=\frac{R^2 + 2 R'^2 - R R''}{(R^2 + R'^2)^{3/2}}
\label{k}
\ee
where $R'$ and $R''$ denote the derivatives with respect to $\theta$.
In figure \ref{fig1} we see that the octupole deformation shows
parts with negative curvature at smaller deformation parameter than the quadrupole deformation shows. However, the Lyapunov exponent of octupole deformation remains 
always smaller. We therefore conclude 
that the octupole deformation compared with the quadrupole deformation 
leads to no 
significant enhancement of chaotization.

\section{Damping of collective oscillations}\label{III}

We would like to focus on a unified description of the linear response including collisional contributions and chaotic scattering with the surfaces. Therefore let us briefly sketch the response function formalism starting from appropriate kinetic equations. For infinite matter this procedure will result in the known Mermin response function which has been used to describe the IVGDR in symmetric \cite{FMW98} and asymmetric nuclear matter \cite{MWF97}. Then we derive the response function for finite nuclei and show that in a local density approximation chaotic scattering from the surface can be incorporated. The result will be a Mermin-like response function in local density approximation where the largest Lyapunov exponent appears as an imaginary shift in the frequency.

\subsection{Infinite matter response}
In order to consider the collisional damping we start
from the kinetic equation for the quasiclassical distribution function. For neutrons this kinetic equation reads
\be
&&{\dot f}_n(p,r,t)+\frac{p}{m}\partial_r f_n(p,r,t)-\partial_r(V_n+V_{\rm ext}) \partial_pf(p,r,t)=I
                      \label{klassVlasov}
\nonumber\\&&
\ee
with the self-consistent mean-field potential given by a schematic Skyrme type\cite{VBR72,BRV94}
\be
V_n(r,t)&=&t_0\left\{\left(1+\frac{x_0}{2}\right)(n_n(r,t)+n_p(r,t))\right
.\nonumber\\
&-&\left . \left(x_0+\frac{1}{2}\right)n_n(r,t)\right\}
\nonumber\\
&+&
\frac{t_3}{4}\left\{(n_n(r,t)+n_p(r,t))^2-n_n^2(r,t)\right\}\label{PotNeu}.
\ee
The neutron and proton densities are $n_n=\int {dp \over (2 \pi \hbar)^3} f_n$ and $n_p=\int {dp \over (2 \pi \hbar)^3} f_p$, respectively. The parameters are $x_0=0.48$, $t_0=-983.4$ MeV fm$^{3}$ and $t_3=13106$ MeV fm$^{6}$. The kinetic equation and mean field for protons are obtained by interchanging the corresponding densities. 
We restrict here to symmetric nuclear matter. Generalizations to asymmetric nuclear matter can be found in \cite{MWF97}. 
First we sketch the results for the collision free ($I=0$) or Vlasov equation and than we will take collisions into account.

\subsubsection{Collision free}

The collective effects without collisions ($I=0$) can be obtained from the linearization of
the Vlasov equation (\ref{klassVlasov}) 
with respect to an external potential $V_{\rm ext}$.
In the case of isovector oscillations, the density variation $\delta n=n_p-n_n$ 
is considered as the difference between proton and neutron densities.
The linearization of the difference of kinetic equations for neutrons (\ref{klassVlasov}) and protons leads to
\be
\delta n=\Pi_0 (\delta V_{\rm ext}+V_0 \delta n)
\label{dn}
\ee
with \cite{BRV94}
\be
V_0=2 (V_n-V_p)&=&-\frac{t_0}{2}\left(x_0+\frac{1}{2}\right)-\frac{t_3}{8} n_0
\label{Potential_V0}
\ee
where $n_0=\int {dp \over (2 \pi \hbar)^3} f_0=0.16$ fm$^{-3}$ is the nuclear saturation density
and $\Pi_0$ is the Lindhard-function
\be
 \Pi_0^{\rm inf}(q,\omega)=
             4\, \int \frac{dp'}{(2\,\pi\hbar)^3} \frac{q\,\partial_{p'}\, 
             f_{0}(p')}{\displaystyle{\frac{p'\,q}{m}}-\omega-i0}.
                                     \label{klassLindT0}
\ee
Here we first consider the homogeneous equilibrium $f_0(p)$ or infinite matter.

The response function $\Pi$ connects the induced density
fluctuation $\delta n$ with the external potential via
\be
\delta n=\Pi\, V_{ext}
\label{polari}
\ee
and the response function follows from (\ref{dn}) for infinite matter
\be
\Pi^{\rm inf}(q,\omega) = 
        \frac{\Pi_0^{\rm inf}(q,\omega)}{1-V_0\, \Pi_0^{\rm inf}(q,\omega)} \equiv 
                          \frac{\Pi_0^{\rm inf}(q,\omega)}{\epsilon(q,\omega)}.
                                           \label{dieletricfunc}
\ee
The zeros $\omega=\Omega+i\gamma$ of the dielectric function  
$\epsilon (q,\omega)=0$ in (\ref{dieletricfunc}) determine the collective modes
with the energy $\Omega$ and the Landau damping $\gamma$ of the
collective excitation. Within the Steinwedel-Jensen model for IVGDR \cite{SJE50}, the wave 
vector scales like
$q=\frac{\pi}{2 R_0}$ where the nuclear radius is $R_0=1.13 A^{1/3}$fm.

\subsubsection{Collisional model}

In order to take collisions into account as a further damping effect beyond Landau damping,
we start from a kinetic equation analogous to (\ref{klassVlasov})
with an additional collisional term $I[p,r,t]$ on the right hand side.
In \cite{FMW98} we have derived a collision integral in a non-Markovian
relaxation time approximation
\be
I(p,r,t)=\int\limits_0^t{{\tilde f}(p,r,{\bar t})-f(p,r,{\bar t}) 
\over \tau(t-{\bar t})} d{\bar t}
\ee
with the dynamical non-Markovian relaxation time
\be
 \frac{1}{\tau_{m}(\omega)}=\frac{1}{\tau_B}\left[1+\frac{3}{4}
 \left(\frac{\omega}{\pi T} \right)^2 \right] \, .\label{taumemory}
\ee
The Markovian relaxation time is given by  
$\tau^{-1}_B=\frac{8\pi m}{3\hbar^3}\sigma T^2$, 
where $\sigma$ is   
the averaged spin-isospin proton-neutron 
cross section. This collision integral holds for low temperatures compared to the Fermi energy. The non-Markovian relaxation time arises from the coupling of collective modes regarding two-particle scattering and consequently describes the effect of zero sound damping. 
The local equilibrium distribution $\tilde f$ is determined through the conservation of the local 
current. The linearization of the kinetic equation leads then to the 
extended response function of Mermin \cite{MER70} 
\be
\Pi^{\rm M}_0(q,\omega)=
{\Pi_0^{\rm inf}(q,\omega+{i\over \tau})\over
1-{i \over \omega\tau +i} (1- {\Pi_0^{\rm inf}(q,\omega+{i\over
\tau}) \over \Pi_0^{\rm inf}(q,0)} )}\label{merminDF},
\ee 
where the selfconsistency leads to the replacement of $\Pi_0^{\rm inf}$ by $\Pi_0^{\rm M}$ in 
(\ref{dieletricfunc}). Generalizations to asymmetric nuclear matter can be found in \cite{MWF97}.

The energy and damping rates are determined by the zeros of the
(Mermin) response function  \cite{FMW98}
\be
\epsilon^M(q,\Omega+i \gamma)=1-V_0(q) \Pi_0^{\rm M}(q,\Omega+i \gamma)=0.
\label{zero}
\ee
Here the damping rate represents Landau and collisional damping.

\subsection{Finite matter response}

In the next step we will present the finite matter response function in terms of a memory integral over all trajectories. It allows us to introduce the local density approximation as a first order memory effect in the trajectories. 
For this purpose we rewrite the Vlasov equation (\ref{klassVlasov}) in a slightly different way.
Introducing the Lagrange picture by following the trajectory $x(t), p(t)$ of 
a particle we linearize the Vlasov 
equation (\ref{klassVlasov}) according to $f(x,p,t)=f_0(x,p)+\delta f(x,p,t)$ as
\be
\frac{d}{dt}\delta f(x(t),p(t),t)= \partial_p f_0\,\partial_{x(t)}\,\delta V
\ee
and get with one integration 
\be
&&\delta f(x,p,t) = \nonumber\\
&&-2m \!\! \int\limits_{-\infty}^0 dt'\!\!\int\limits_{-\infty}^{\infty} 
                    \!\!dx'\frac{d}{dt'}\delta(x'-x(t')) \; 
                {\partial f_0(p^2,x') \over \partial {p^2}}\,\delta V(x',t+t').  
\nonumber \\
\ee
The density variation caused by the external potential is obtained as
\be
\delta n(x,\omega)&=& 
-2ms\int dx'\int  \frac{dp^3}{(2\pi\hbar)^3}\partial_{p^2}
                                         f_0(p^2,x')\nonumber\\
&&\times\int\limits_{-\infty}^0 dt'e^{-it'
      \omega}\frac{d}{dt'}\delta(x'-x(t'))\delta V(x',\omega) 
\ee
where $s$ denotes the spin-isospin degeneracy. Comparing this expression with 
the definition of the polarization function 
$\Pi_0$,
\be
\delta n(x,\omega)=\int dx'\,\Pi_0(x,x',\omega)\;\delta V(x',\omega),
\ee
we are able to identify the polarization of finite systems as
\be
\Pi_0(x,x',\omega)&=&- 2ms\int \frac{dp^3}{(2\,\pi\,\hbar)^3}\partial_{p^2}\,
f_0(p^2,x')\nonumber\\
&\times&\int\limits_{-\infty}^0 dt'\, e^{-i\,t'\,
\omega}\frac{d}{dt'}\delta(x'-x(t')). 
\ee

Further simplifications are possible if we focus on the ground state
$f_0(p^2)=\Theta(p_f^2-p^2)$. The modulus integration of momentum can 
be carried out and the Kirzhnitz-formula 
\cite{kirschnitz,DM95} for the polarization function appears
\be
&&\hspace{-.7cm}\Pi_0(x,x',\omega)= \nonumber\\
&&-\frac{m s p_f(x)}{4 \pi^2 \hbar^3}\left[ 
\int\limits_{-\infty}^0 dt'  e^{-i t' \omega}
\frac{d}{dt'}\int\frac{d\Omega_p}{4 \pi}\delta(x'-x(t')) \right] \nonumber\\
&=& -\frac{m s p_f(x)}{4 \pi^2 \hbar^3}\Bigg[ 
\delta(x'-x(0))\nonumber\\
&&+\,i  \omega\int\limits_{-\infty}^0 dt'  e^{-i t' 
\omega}\int\limits_{}^{} 
\frac{d\Omega_p}{4 \pi}\delta(x'-x(t'))\Bigg]. 
\label{prewigner}
\ee
This formula represents the ideal free part and a contribution which arises 
by the trajectories $x(t)$ averaged over the 
direction at the present time ${\vec n}_p p_f=m {\dot {\vec x}}(0)$. In principle, the knowledge of the evolution of all 
trajectories is necessary to 
evaluate this formula. Avoiding the latter expense, we discuss two approximations which will give us an insight into the physical processes behind. 
First the most restrictive one shows how the local 
density approximation emerges. As an extension we consider then the influence of 
chaotic scattering on a surface.

\subsubsection{Local density approximation}
The local density approximation appears from (\ref{prewigner}) when two simplifications are performed. Introducing Wigner 
coordinates $R=(x+x')/2$, $r=x-x'$ one has to assume
\begin{enumerate}

\item
gradient expansion
\be
p_f(R+\frac{r}{2})\approx p_f(R) +{\cal O}(\partial_R)
\ee
\item
expansion of the trajectories up to first order history 
\be
x'-x(t')\approx -r-t' {\dot x} =
          -r-t' \frac{p_f}{m}  {\vec n}_p + {\cal O}(t'^2).
\label{rule}
\ee
\end{enumerate}
With these two assumptions we obtain from (\ref{prewigner}) after trivial 
integrations
\be
&&\Pi_0^{\rm LDA}(q,R,\omega)= \nonumber\\
&&-\frac{m s p_f(R)}{4 \pi^2 \hbar^3} 
\Bigg\{
1+i \zeta \int\limits_{0}^{\infty} dy  
{\rm e}^{i\zeta y}\frac{\sin \displaystyle{y}}{\displaystyle{y}}\Bigg\}
                        \label{pre1}
\ee
where ${\zeta}={m\omega\over q p_f(R)}$. This can be further integrated with the help of
\be
&&\int\limits_0^{\infty} dy {\rm e}^{i \zeta y} {\sin y\over y}= {\rm arctan} ({\rm Im} \, \zeta-i {\rm Re} \, \zeta)^{-1}\nonumber\\
&& =\left . 2 i\ln \left( 
\frac{1+\zeta} {1-\zeta} \right) + \pi \left[\mbox{sgn}\left(1+\zeta
\right)+\mbox{sgn}\left(1-\zeta \right) \right]\right |_{{\rm Im}\, \zeta \to 0}\nonumber\\
&&
\ee
leading to the standard 
Lindhard result for (\ref{pre1}).
We recognize the ground state result for infinite 
matter (\ref{klassLindT0}) except that the Fermi 
momentum $p_f(R)$ has to be understood as a local 
quantity with respect to the density 
\be
\Pi_0^{\rm LDA}(q,R,\omega)=\Pi_0^{\rm inf}(q,p_f(R),\omega)\label{lda}.
\ee
For extensions beyond the local density approximation see \cite{DM95,DM90}.

\subsection{Influence of chaotic scattering with surface on damping}

Now we want to assume an additional chaotic scattering
which will be caused e.g. by the curved surface. In order 
to investigate this effect we add to the regular motion (\ref{rule}) a small 
irregular part $\Delta x$ 
\be
x'-x(t')\approx -r-t' \frac{p_f}{m}  {\vec n}_p +\Delta x.
\label{rule2}
\ee
This irregular part of the motion has the direction of the velocity, ${\vec n}_p$, 
and lasts a time $\Delta_t$. During this time
an exponential increase in phase-space occurs controlled by the largest 
Lyapunov exponent 
$\lambda$. Therefore we assume [$t'<0$]
\be
\Delta_x\approx {p_f {\vec n_p} \over m} \Delta_t \exp[-\lambda 
(t'-\Delta_t)]+\mbox{const.} \label{deltax}
\ee
Since we are looking for the upper bound of Lyapunov exponent 
we can take (\ref{deltax}) 
at the maximum $\Delta_t=-1/\lambda$. 
Furthermore, in the case of vanishing Lyapunov exponent, the regular motion (\ref{rule}) should be recovered. This determines the constant. We 
obtain finally
\be
x'-x(t')\approx -r-\frac{p_f}{m}  
{\vec n}_p \left [{1-\exp(-\lambda t') \over \lambda} \right ].
\label{rule3}
\ee
With this ansatz one derives from (\ref{prewigner}) the result
\be
\Pi(q,R,\omega)&=&-\frac{m s p_f(R)}{4 \pi^2 \hbar^3}\Bigg[\nonumber\\
&&1+i \zeta\int\limits_{0}^{\infty} d y \frac{\sin 
\displaystyle{y}}{\displaystyle{y}} \left (1+ {\zeta y \over \omega} \lambda\right )^{i \omega/\lambda-1}\Bigg],\nonumber\\
&&                                                       \label{pre2}
\ee
which for $\lambda\rightarrow 0$ resembles exactly (\ref{pre1}). The further 
integration could be given in terms of 
hypergeometric functions but this is omitted here.

With formula (\ref{pre2}) we have derived the main result of this section, i.e. a polarization function due to many 
particle effects including the influence of an additional chaotic process 
characterized by the Lyapunov exponent $\lambda$.

For small values of ${m\over q p_f} \lambda=\zeta {\lambda \over \omega}$ which corresponds to relative small Lyapunov exponents, we can 
use $\lim\limits_{x \to \infty}(1+a/x)^x=\exp(a)$ in the integral of (\ref{pre2}) and the final integration can be
performed with the result of (\ref{lda}) and a complex shift in the frequency
\be
\Pi_0^{\rm surf}(q,R,\omega)=\Pi_0^{\rm inf}(q,p_f(R),\omega+i \lambda)
\label{ldac}.
\ee

Solving the dispersion relation we obtain in this way the Matthiessen rule which states that different 
damping mechanisms are additive in the total
damping. However, the collisional damping of infinite matter, $2 \gamma$, and the largest Lyapunov exponent, $\lambda$, are not yet the appropriate values of finite matter. Instead, these two damping mechanisms have to be added by a proper relative weight which will take into account the relative occurrence of the processes, the collisions with other particles and the collisions with the surface.

\section{Comparison of damping mechanisms}\label{IV}

\subsection{Connection between surface curvature and temperature}

The damping of excited nuclei can be understood by two different
mechanisms. Besides the collisional damping we also have to consider the shape fluctuations. In the absence of inertia which will be assumed in the following, the driving force for shape fluctuations is the temperature. Therefore we link the surface 
deformation $\alpha$ of (\ref{r1}) to the 
temperature within a statistical model. We use as a measure for the mean deformation \footnote{For small deviations we found identical Lyapunov exponents for prolate $\alpha>0$ and oblate $\alpha<0$ deformations and therefore we do not distinguish the sign of $\alpha$.}
\be
\langle\alpha\rangle={\int d \alpha |\alpha| \exp{(-E_B(\alpha)/T)} \over 
\int d \alpha \exp{(-E_B(\alpha)/T)}}
\label{38}
\ee
where the surface dependent energy $E_B(\alpha)$ is given by the Bethe-Weizs\"acker 
formula \footnote{Please remind that in principle the Coulomb energy changes with small deformation as well according to the factor \protect\cite{greinermaruhn}
\be
1-5 {(\lambda-1) \over (2 \lambda+1)^2} \alpha_{\lambda}^2
\label{35}
\ee 
while the surface term changes as 
\be
1+(\lambda-1)(\lambda+2)/2/(2\lambda+1) \alpha_{\lambda}^2.
\label{factor}
\ee
Only the correction (\ref{factor}) is considered since (\ref{35}) would lead to corrections of around $0.3\%$ and are neglected here.}
\be
&&
{E_B}(\alpha)= -a_1 +{a_2\over A^{1/3}}+{a_3 Z^2\over A^{4/3}} + a_4 
\delta^2+ a_5 A^{2/3} {S(\alpha) \over S(0)}\nonumber\\&&
\ee
with the volume energy $a_1=15.68$ MeV, Coulomb energy $a_3=0.717$ MeV, 
the symmetry energy $a_4=28.1$ MeV and the surface 
energy $a_5=18.56$ MeV.

The surface of rotational symmetric nuclei (\ref{r1}) is given by
\be
S=2 \pi \int\limits_{-1}^1 dx R_\lambda(x) \sqrt{R_\lambda(x)^2+(1-x^2) R_\lambda'(x)^2}
\ee
from which we obtain for the quadrupole, $S_2$, and octupole, $S_3$, deformations (\protect\ref{factor})
\be
{S_2(\alpha) \over S(0)}&=&1+\frac 2 5 \alpha^2+
                                                   {\cal O}(\alpha^3)\nonumber\\
{S_3(\alpha) \over S(0)}&=&1+\frac 5 7 \alpha^2+
                                                   {\cal O}(\alpha^3). \label{c}
\ee
This represents the lowest order expansion in $\alpha$, 
however the next term gives already corrections in fractions of percents
for the highest deformations considered here.
By this way, the statistical model (\ref{38}) leads to a connection between temperature and 
deformation as
\be
T=c \pi a_5 A^{2/3} <\alpha>^2\label{T}
\ee
where the constant $c$ is given by the coefficient of $\alpha^2$
in Eq. (\ref{c}) for the corresponding quadrupole or octupole deformation.

\subsection{Finite matter scaling for collisional damping}

According to the local density approximation (\ref{lda}) we want to investigate the 
collisional damping in finite matter. Therefore we have to replace all densities
in the dispersion relation (\ref{merminDF}) 
 by the local 
density which are parameterized by a Woods-Saxon potential
\be
n(r)=\frac{n_0} {\exp\left[(r-R_0)/0.545\,{\rm fm}\right]+1}.
\ee
Solving the dispersion relation (\ref{zero}) within the local density approximation according to (\ref{lda}) we obtain a spatial dependent damping rate $\gamma(r)$.

In the following we derive a finite size scaling factor, $\xi_A$, of the damping 
\be
\xi_A=\frac{\langle\gamma(r)\rangle}{\gamma}, \label{scaling}
\ee
which 
approximates the local ($r$-dependent) dispersion 
relation by an averaged one
\be
\langle \ldots \rangle ={3 \over R_0^3} \int_0^{R_0} r^2\ldots\,d\,r
\ee
assuming radial symmetry. The factor (\ref {scaling}) has to be applied to the infinite matter bulk collisional damping and
is found $\xi_{\rm Sn}=0.9199$ and $\xi_{\rm Pb}=0.9250$ for $^{120}$Sn and 
$^{208}$Pb respectively. Therefore, the bulk matter value of damping is diminished by finite size effects.

\subsection{Comparison of collisional damping and surface damping}

\begin{figure}[t]
  \centerline{\psfig{file=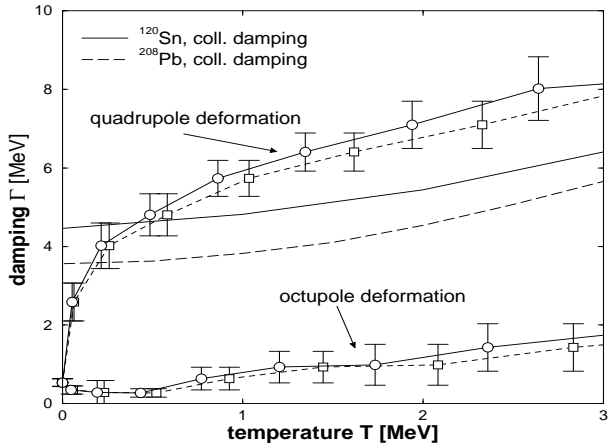,width=8cm,height=6cm,angle=-90}}
  \caption{The collisional damping scaled with finite size according to (\protect\ref{scaling}) is
           compared with the damping according to the 
           chaotic scattering from the surface of quadrupole and 
           octupole deformed shapes for $^{120}$Sn and $^{208}$Pb from figure \protect\ref{fig1}.} 
                                                              \label{fig3}
\end{figure}

Using Eq. (\ref{T}) we can translate 
the Lyapunov exponent $\lambda$ calculated as a function of deformation in figure \ref{fig1} into a function of the temperature. 
In figure~\ref{fig3} the contribution to 
the damping of IVGDR for $^{120}$Sn (circles) and $^{208}$Pb (squares) 
is presented for different shape deformations versus temperature.
If we add quadrupole and octupole deformations 
we come up with a damping curve very 
similar to the paper of Ormand \cite{OBB97}. The damping starts at zero and 
increases rapidly with increasing temperature. We see 
that the main contribution comes from the quadrupole deformation while 
the octupole deformation is only sizeable at higher 
temperature. Let us note that the qualitative difference between 
Sn and Pb is reproduced by surface scattering as well as collisional damping.

\begin{figure}[t]
  \centerline{\psfig{file=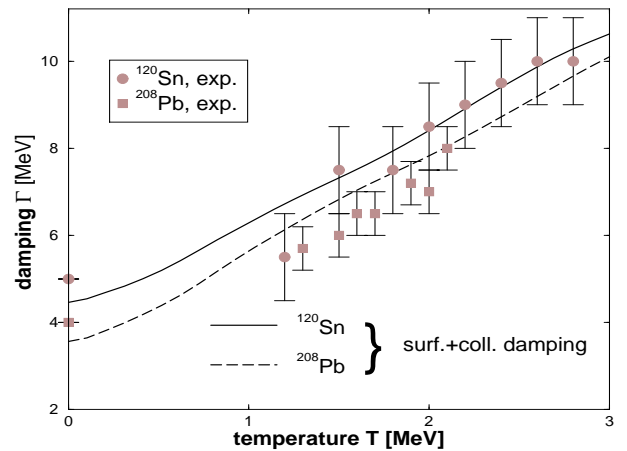,width=8cm,height=6cm,angle=-90}}
  \caption{The effective damping consisting of collisional and surface 
damping together with the experimental data (filled symbols)
           from Ref. \protect\cite{RAP96} (Sn) and from 
           Ref. \protect\cite{RAM96} (Pb). }        \label{fig4}
\end{figure}

The collisional contribution $2 \gamma$ scaled 
to finite sizes via (\ref{scaling}) are 
plotted as well (Sn: solid line, Pb: dashed line). We recognize that
both contributions by itself, collisional as well as surface scattering, account 
almost for the same amount required by the experimental values, see figure \ref{fig4}. A proper relative weight between both processes is therefore necessary which will be introduced in the following.

So far we have not considered that only particles close
to the surface can appreciably contribute to the
surface chaotization, while particles deep inside
the nuclei are screened out of this process.
Consequently we consider the corresponding collision frequencies as the measure to compare surface collisions with interparticle collisions. The collision frequency between particles is given by $1/\tau_M$ of (\ref{taumemory}). The collision frequency of particles with the deformed surface beyond a sphere, $\nu_{\rm surf}$, is given 
by the product of the density with the surface increase
$S(\alpha)-S(0)=c \alpha^2 4 \pi R_0^2$ according to Eq. (\ref{c}) and with
the mean velocity in radial direction $v_r=3/8 v_F$. The result is
\be
\nu_{\rm surf}=1.5 T n_0 v_F r_0^2/a_5
\label{surf}
\ee
where we have used Eq. (\ref{T}) to replace $\alpha$.
We see that the frequency (\ref{surf}) is independent of the size of the nucleus and 
linearly dependent on the temperature.

We use the ratio of these two frequencies to weight properly the two damping mechanisms, the surface collisional, $\lambda$, and interparticle collisional, $<\gamma(r)>$, contributions. Consequently the full - width - half - maximum reads (FWHM)
\be
\Gamma_{\rm FWHM}&=&2 \left (\zeta \, <\gamma(r)>+(1-\zeta) \, \lambda \right )\nonumber\\
&\equiv&\Gamma_{\rm coll}+\Gamma_{\rm surf}.
\label{eff}
\ee
With the help of (\ref{taumemory}) and (\ref{surf}) the weighting factor $\zeta$ is given by
\be
 &&\zeta(T)={{1\over \tau_M(T)}\over {1\over \tau_M(T)}+\nu_{\rm surf}(T)}.
\ee
One sees that for zero and high temperatures $\zeta=1$ and 
due to Eq. (\ref{eff}) only the collisional contributions matter. 
Since $\nu_{\rm surf}$ is linear in the temperature and $1/\tau$ depends quadratically on the temperature, the weighting factor $\zeta$ has a minimum at temperatures around $T_c={\sqrt{3} \over 2 \pi}\omega $  and the surface contributions become important. In the case of the IVGDR this corresponds to a temperature of $T\approx 3.7$MeV, which is the upper limit of current experimental achievable temperatures. Therefore we can state that at low and high temperatures the collisional damping is dominant while for temperatures around $T_c$ the surface contribution becomes significant.

In figure \ref{fig4} we compare the effective damping  
according to Eq. (\ref{eff})
with the experimental data. We find a reasonable quantitative agreement. 
While the collisional damping by itself can reproduce the increase at 
small temperatures well enough it fails to reproduce the sharp rise at higher temperatures, see figure~\ref{fig3}. 
The surface damping by itself fails at zero temperature and leads to an unphysical shape.
Only the connection of both effects, collisional and surface damping 
seems to be able to reproduce the data correctly.

At this point we like to stress that we have adopted a simplified model of finite Fermi liquids and introduced a phenomenological weighting between surface and interparticle collision contributions proportional to their relative occurrence. The discussion of damping due to finite size effects within shell model considerations can be found in literature, e.g. the review \cite{BBB83} where this problem has been addressed from the point of doorway states.

\section{Summary}\label{V}

We have considered the influence of collisional damping as well as surface scattering for 
the damping of hot isovector giant dipole resonances (IVGDR). We find that the relative importance of 
both effects is dependent on the temperature scale. While for low and high temperatures the 
collisional contribution dominates, the surface scattering is the one of significant importance at temperatures around $28\%$ of the centroid energy.
 
The surface contribution has been calculated from the Lyapunov 
exponent of surface scattering for different deviations of the nucleus shape. The deviations 
are related to the temperature by a statistical model.

The influence of both the surface and collisional damping is described by a generalized response function for 
finite nuclear matter which takes into account the chaotic processes of 
scattering with the surface. We derive by this way a response function similar to the 
Lindhard response in local density approximation which is now modified by the Lyapunov 
exponent of the surface scattering. In an approximative way we show that 
the total damping is described by the sum of the two contributions to the damping 
in analogy to the 
Matthiessen rule. 

Comparing the collision frequencies for one particle with other particles and 
for one particle with the surface, we derive a proper relative weight for both processes, 
the collisional and the surface damping. The resulting effective damping reproduces the 
experimental data rather well in the whole accessible range.

\acknowledgements

The authors are especially indebted to A. Dellafiore for bringing the Kirzhnitz 
formula to our attention. M. DiToro is thanked for many discussions and Hans J. Weber for helpful comments.
The work was supported by
the BMBF (Germany) under contract
Nr. 06R0884, the DFG under contract Nr. 905-13/1 and the Max-Planck Society.
 

\begin{thebibliography}{10}

\bibitem{BBB83}
G. Bertsch, P. Bortignon, and R. Broglia, Rev. Mod. Phys. {\bf 55},  287
  (1983).

\bibitem{KAM69}
S. Kamerdzhiev, Yad. Fiz. {\bf 9},  324  (1969).

\bibitem{SAB84}
H. Sagawa and G.~F. Bertsch, Phys. Lett. B {\bf 5},  138  (1984).

\bibitem{ATS94}
V. Abrosimov, M. Di~Toro, and A. Smerzi, Z. Phys. A {\bf 347},  161  (1994).

\bibitem{KPS95}
V. Kolomietz, V. Plujko, and S. Shlomo, Phys. Rev. C {\bf 52},  2480  (1995).

\bibitem{KPS96}
V. Kolomietz, V. Plujko, and S. Shlomo, Phys. Rev. C {\bf 54},  3014  (1996).

\bibitem{KTO96}
V. Kondratyev and M. Di~Toro, Phys. Rev. C {\bf 53},  2176  (1996).

\bibitem{HNP96}
E. Hern\'andez, J. Navarro, A.Polls, and J. Ventura, Nucl. Phys. {\bf A597},  1
   (1996).

\bibitem{TKL97}
M. Di~Toro, V. Kolomietz, and A. Larionov, in {\em Proceedings of the Dubna
  Conference on Heavy Ions}, Dubna, 1997 (unpublished).

\bibitem{KLT97}
V. Kolomietz, A. Larionov, and M. Di~Toro, Nucl. Phys. {\bf A613},  1  (1997).

\bibitem{FMW98}
U. Fuhrmann, K. Morawetz, and R. Walke, Phys. Rev. C. {\bf 58},  1473  (1998).

\bibitem{KLP98}
V.~M. Kolomietz, S.~V. Lukyanov, V.~A. Plujko, and S. Shlomo, Phys. Rev. C {\bf
  58},  198  (1998).

\bibitem{MWF97}
K. Morawetz, R. Walke, and U. Fuhrmann, Phys. Rev. C {\bf 57},  {R 2813}
  (1998).

\bibitem{AYG98}
S. Ayik, O. Yilmaz, A. Golkalp, and P. Schuck, Phys. Rev. C {\bf 58},  1594
  (1998).

\bibitem{GTO98}
G. Gervais, M. Thoennessen, and W.~E. Ormand, Phys. Rev. C {\bf 58}, 1377 (1998).

\bibitem{SHB75}
S. Shlomo and G.~F. Bertsch, Nucl. Phys. {\bf A243},  507  (1975).

\bibitem{ESB83}
H. Esbensen and G.~F. Bertsch, Phys. Rev. C {\bf 28},  355  (1983).

\bibitem{ESB84}
H. Esbensen and G.~F. Bertsch, Ann. Phys. {\bf 157},  255  (1984).

\bibitem{BBB91}
P.~F. Bortignon, A. Bracco, D. Brink, and R.~A. Broglia, Phys. Rev. Lett. {\bf
  67},  3360  (1991).

\bibitem{OBB96}
W.~E. Ormand, P.~F. Bortignon, and R.~A. Broglia, Phys. Rev. Lett. {\bf 77},
  607  (1996).

\bibitem{KST97}
S. Kamerdzhiev, J. Speth, and G. Tertychny, Nucl. Phys. {\bf A624},  328
  (1997).

\bibitem{OBB97}
W.~E. Ormand, P.~F. Bortignon, R.~A. Broglia, and A. Bracco, Nucl. Phys {\bf
  A614},  217  (1997).

\bibitem{KLL98}
S. Kamerdzhiev, R.~J. Liotta, E. Litvinova, and V. Tselyaev, Phys. Rev. C {\bf
  58},  172  (1998).

\bibitem{KIN98}
J. Kvasil, N.~L. Iudice, V.~O. Nesterenko, and M. Kopal, Phys. Rev. C {\bf 58},
   209  (1998).

\bibitem{ALB89}
Y. Alhassid and B. Bush, Phys. Rev. Lett. {\bf 63},  2452  (1989).

\bibitem{OBB90}
W.~E. Ormand {\it et~al.}, Phys. Rev. Lett. {\bf 64},  2254  (1990).

\bibitem{OCB92}
W.~E. Ormand {\it et~al.}, Phys. Rev. Lett. {\bf 69},  2905  (1992).

\bibitem{LBB95}
B. Lauritzen, P.~F. Bortignon, R.~A. Broglia, and V.~G. Zelevinsky, Phys. Rev.
  Lett. {\bf 74},  5190  (1995).

\bibitem{BCM95}
A. Bracco {\it et~al.}, Phys. Rev. Lett. {\bf 74},  3748  (1995).

\bibitem{LDH96}
S. Leoni, T. D{\o}ssing, and B. Herskind, Phys. Rev. Lett. {\bf 76},  4484
  (1996).

\bibitem{CTG93}
M. Colonna {\it et~al.}, Phys. Lett. B {\bf 307},  293  (1993).

\bibitem{BAR96}
V. Baran {\it et~al.}, Nucl. Phys. {\bf A599},  29  (1996).

\bibitem{MWS98}
R. Walke and K. Morawetz, Phys. Rev. C. {\bf 60},  17301  (1999).

\bibitem{SLM96}
V. {\v S}pi{\v c}ka, P. Lipavsk{\'y}, and K. Morawetz, Phys. Lett. A {\bf 240},
   160  (1998).

\bibitem{MLSCN98}
K. Morawetz {\it et~al.}, Phys. Rev. Lett. {\bf 82},  3767  (1999).

\bibitem{BDT86}
D. Brink, A. Dellafiore, and M. DiToro, Nucl. Phys. A {\bf 456},  205  (1986).

\bibitem{MJ36}
N.~F. Mott and H. Jones, {\em The Theory of the Properties of Metals and
  Alloys} (Oxford University Press, London, 1936).

\bibitem{swiatecki}
J. Blocki, J.-J. Shi, and W. Swiatecki, Nucl.Phys {\bf A554},  387  (1993).

\bibitem{ringschuck}
P. Ring and P. Schuck, {\em The Nuclear Many-Body Problem} (Springer-Verlag,
  New York, 1980).

\bibitem{VBR72}
D. Vautherin and D. Brink, Phys. Rev. C {\bf 5},  626  (1972).

\bibitem{BRV94}
F. Braghin and D. Vautherin, Phys. Lett. B {\bf 333},  289  (1994).

\bibitem{SJE50}
H. Steinwedel and J. Jensen, Z. Naturforsch. {\bf 5},  413  (1950).

\bibitem{MER70}
N. Mermin, Phys. Rev. B {\bf 1},  2362  (1970).

\bibitem{kirschnitz}
D. Kirzhnitz, Y. Lozovik, and G. Shpatakovskaya, Usp. Fiz. Nauk {\bf 117},  3
  (1975).

\bibitem{DM95}
A. Dellafiore, F. Matera, and D.~M. Brink, Phys. Rev. A {\bf 51},  914  (1995).

\bibitem{DM90}
A. Dellafiore and F. Matera, Phys. Rev. A {\bf 41},  4958  (1990).

\bibitem{greinermaruhn}
W. Greiner and J. Maruhn, {\em Nuclear Models} (Springer-Verlag, Berlin et.al,
  1996).

\bibitem{RAP96}
E. Ramakrishnan {\it et~al.}, Phys. Rev. Lett. {\bf 76},  2025  (1996).

\bibitem{RAM96}
E. Ramakrishnan {\it et~al.}, Nucl. Phys. {\bf A549},  49  (1996).

\end{thebibliography}

\end{document}